\documentclass{ws-ijmpa}

\begin{document}

\markboth{Raphael Granier de Cassagnac}
{What's the matter at RHIC?}

%
\catchline{}{}{}{}{}
%

\title{WHAT'S THE MATTER AT RHIC?}

\author{RAPHAEL GRANIER DE CASSAGNAC}

\address{Laboratoire Leprince-Ringuet, \'Ecole polytechnique/IN2P3, 91128 Palaiseau, France.\\
raphael@in2p3.fr}

\maketitle

\begin{history}
\received{Day Month Year}
\revised{Day Month Year}
\end{history}

\begin{abstract}

I present here a concise review of the experimental results obtained at the Relativistic Heavy Ion Collider (RHIC), which shed light on the hot and dense quark gluon matter produced at these high temperature and density conditions.


\end{abstract}

\ccode{PACS numbers: 11.25.Hf, 123.1K}

\section{Foreword}	

This review has little to do with the topic of the symposium it was given in, namely super-symmetry at the Large Hadron Collider (LHC). Indeed, it deals with standard QCD at high density and temperature and how it was widely probed at the Relativistic Heavy Ion Collider (RHIC). In fact, an intimate theoretical link does exist, as first stated by Maldacena\cite{Maldacena:1997re}, between strongly coupled four dimensions super-symmetric\footnote{But is QCD super-symmetric?} Yang Mills theories and weakly coupled type IIb string theories on Anti-deSitter five dimensions space. This so-called AdS/CFT correspondence allows one to compute some properties (viscosity/entropy ratio, quenching parameters...) of the quark gluon matter, as if it were a black hole. Even if such a connection exists on the theory side, I will not comment further on this matter and rather review various experimental facts that highlight certain properties of the matter produced in relativistic (up to $\sqrt{s_{NN}}=200$~GeV per pair of nucleons) heavy ion collisions. The early and most striking RHIC results were summarized in 2005 by the BRAHMS, PHENIX, PHOBOS and STAR experiments in their so-called \emph{white papers}\cite{Arsene:2004fa,Adcox:2004mh,Back:2004je,Adams:2005dq} that will be largely referenced thereafter.

\section{Multiplicities and densities}

\begin{figure}[htpb]
\centerline{\psfig{file=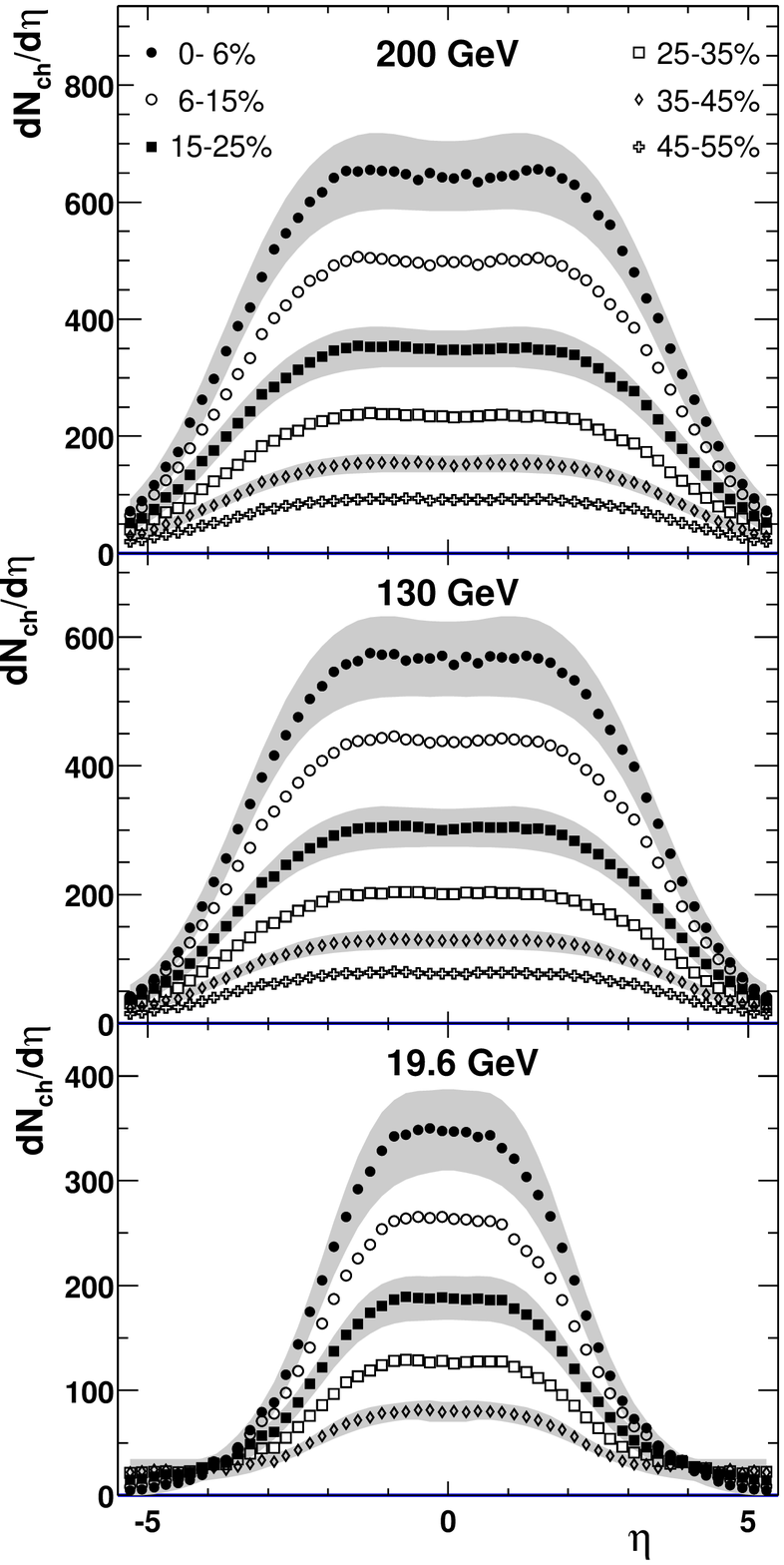,height=6.8cm} \hspace{2em} \psfig{file=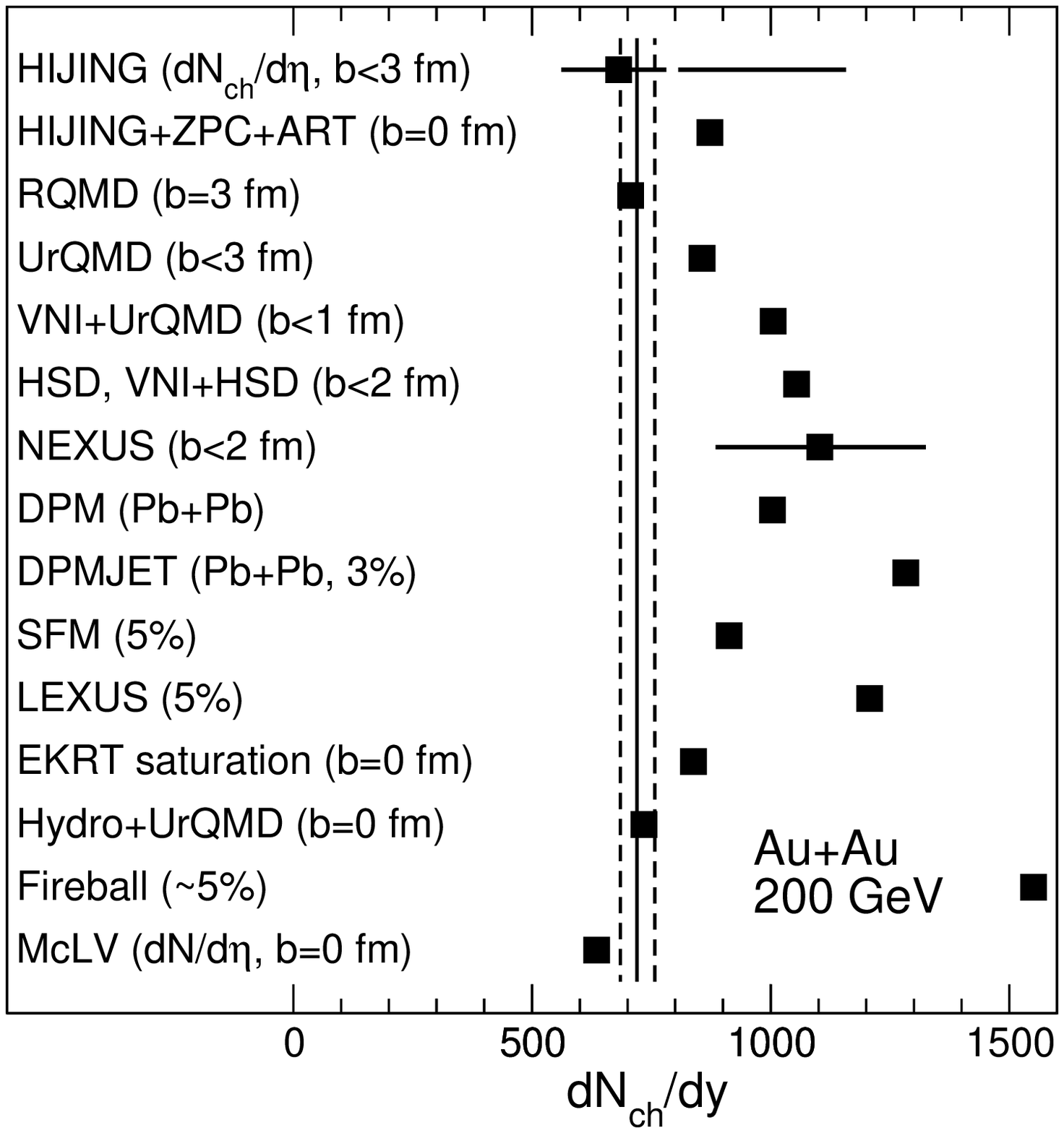,height=6.8cm}}
\vspace*{8pt}
\caption{Left: Charged particle multiplicities as a function of pseudorapidity for various energy and centrality Au+Au collisions from the PHOBOS experiment. Right: The top most $dN_{ch}/dy|_{y=0}$ compared to various models.}
\label{f0}
\end{figure}

The first obvious thing that comes out of a heavy ion collisions is \emph{a lot} of particles. The number of charged particles was measured for various collision energies and centralities by the four RHIC experiments, and in particular by the dedicated PHOBOS collaboration over a broad range of 10.8 units of pseudorapidity\cite{Back:2002wb} (as illustrated by Fig.~\ref{f0} on the left). At midrapidity, the number of charged particle reaches $dN_{ch}/d\eta|_{\eta=0} \simeq 670$ in the most violent Au+Au collisions and they sum up to about 6000 particles (of any charge) over the full rapidity range. As illustrated by the right part of Fig.~\ref{f0}, these huge numbers were in fact \emph{lower} than expected from various simple models, extrapolating lower energy results (for more details and complete references, see section~2.1 in Ref.~\refcite{Back:2004je} or Ref.~\refcite{Eskola:2001vs}). This moderation of the produced particles is an indication that the gluon density in the initial state starts to saturate, or similarly to be shadowed. In other words, low momentum gluons from neighbor nucleons overlap and recombine. Indeed, a strong gluon shadowing in the first model on the figure (HIJING) is necessary to reduce the multiplicities to the observed level. Such a saturated initial state is also described in the Color Glass Condensate (CGC) framework, which is fairly able to reproduce the particle multiplicities (the McLerran-Venugopalan ``McLV'' last model is such a saturation model\cite{Krasnitz:2000gz}).

Another glimpse into the CGC is provided by the high transverse momentum ($p_T$) particle suppression observed in d+Au collision at the highest (up to $\eta = 3.2$) pseudorapidity by the BRAHMS experiment\cite{Arsene:2004ux}. Indeed, higher rapidity allows one to probe lower gluon momentum fractions $x$ in the gold nucleus and one clearly sees an increasing suppression with rapidity (see section~7 in Ref.~\refcite{Arsene:2004fa} and references therein).


These results show that \textbf{the (initial) matter is gluon saturated}.

The second obvious manifestation of the collision violence is the transverse (i.e. unboosted by the initial parton longitudinal momenta) energy liberated. Measuring it allows one to estimate the energy density $\varepsilon$ of the medium after a given time $\tau_0$, through the Bjorken formula\cite{Bjorken:1983}: $\varepsilon = dE_T/dy|_{y=0} / \tau_0 A_T $, where $A_T$ is the transverse area of the collision. The four RHIC experiments measure consistent values of $dE_T/dy|_{y=0}$ that correspond to an energy density of at least 5~GeV/fm$^3$ at $\tau_0=1$~fm/$c$. The question of the time to be considered is not trivial, but 1~fm/$c$ is a maximum if one cares about the earliest as possible thermalized medium. Indeed, hydrodynamics models provide thermalization times between 0.6 and 1~fm/$c$, while the formation time is estimated to be 0.35~fm/$c$ and the nucleus-nucleus crossing time is 0.13~fm/$c$. For a detailed discussion of energy density and time scale estimates, see section~2 of Ref.~\refcite{Adcox:2004mh}. What matters here is that the \emph{lower} energy density estimate is much higher than the threshold for the transition to a quark gluon plasma, as predicted by QCD on the lattice\cite{Karsch:2001cy}: $\varepsilon_c \sim 1$~GeV/fm$^3$.


This tells us that \textbf{the matter should be deconfined}, i.e. made of free quarks and gluons. The following sections review the main measurements that indicate that it is indeed the case.

\section{High Transverse Momentum Suppression}

\begin{figure}[htpb]
\centerline{\psfig{file=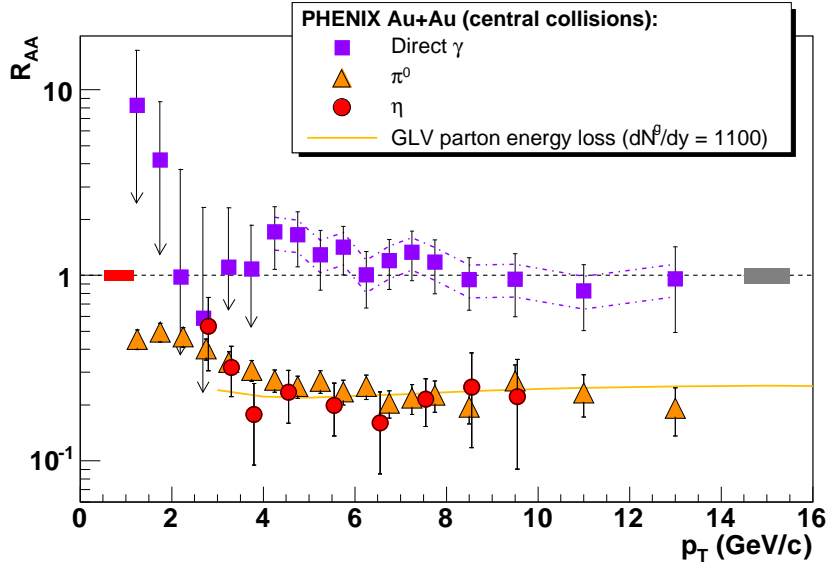,width=11cm}}
\vspace*{8pt}
\caption{Nuclear modification factors for photon, $\eta$ and $\pi^0$ for central collisions, from the PHENIX experiment .}
\label{f1}
\end{figure}

Fig.~\ref{f1} is an illustration of the first and most striking QGP signature seen at RHIC, namely the quenching of jets\cite{Adler:2003qi,Adler:2006bv}. Displayed is the nuclear modification factor $R_{AA}$ defined as the yield of particles seen in A+A collisions, normalized by the same yield from p+p collisions scaled by the average number of binary collisions corresponding to the considered centrality: $R_{AA} = dN_{AA} / \langle N_{coll} \rangle \times dN_{pp}$. Hard processes (high $p_T$ particles in particular) are expected to respect such a scaling ($R_{AA}=1$). This is indeed the case of the direct photon\cite{Adler:2005ig} (purple squares) up to 13~GeV/c, while the corresponding $\pi^0$ (orange triangles) and $\eta$ up to 10~GeV/c (red circles) are suppressed by a factor of five\footnote{It is to be noted that PHENIX has released preliminary data up to 20~GeV/c for $\pi^0$ and 18~GeV/c for photons\cite{Isobe:2007ku}. While the $\pi^0$ remain at $R_{AA}\sim 0.2$, photons start to deviate below unity, possibly because of the nucleus to proton isospin difference\cite{Arleo:2006xb}, which has nothing to do with QGP.}. This is understood as an energy loss of the scattered partons going through a very dense matter, and producing softened jets and leading (high $p_T$) particles. The medium is so dense that it cannot be made of individual hadrons, but rather of quarks and gluons. Gluon densities of the order of $dN^g/dy \sim 1100$ are needed to produce such a strong quenching\cite{Vitev:2002pf}.


High $p_T$ suppression is seen for various particles with various $p_T$ reaches and by the four experiments\cite{Arsene:2004fa,Adcox:2004mh,Back:2004je,Adams:2005dq}. It gets stronger for more central collisions. It is not observed in d+Au collisions (in particular for neutral pions\cite{Adler:2003ii} to be compared to the ones on Fig.~\ref{f1}) where a moderate enhancement is even seen as a function of $p_T$, probably due to multiple scattering of the incoming partons providing additional transverse momentum (the so-called Cronin effect).

This quenching of high $p_T$ particles shows that \textbf{the matter they traverse is dense}.

\section{Back to Back Jets}

\begin{figure}[htpb]
\centerline{\psfig{file=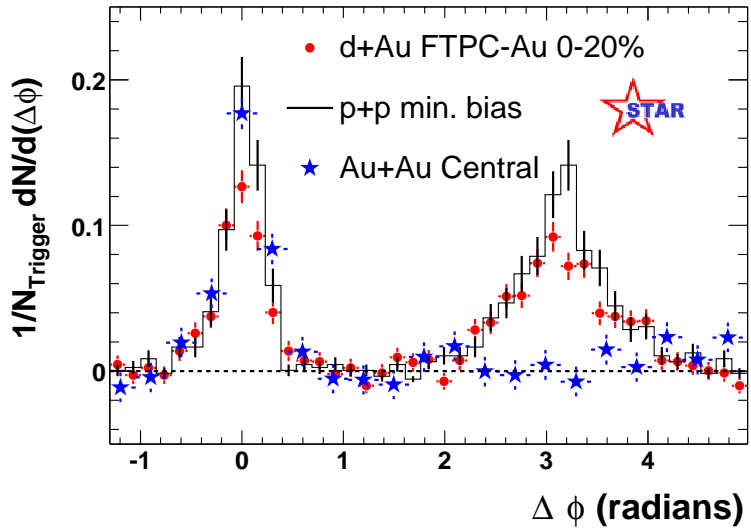,width=11cm}}
\vspace*{8pt}
\caption{Dihadron azimuthal correlations in p+p, d+Au and Au+Au central collisions, from the STAR experiment.}
\label{f2}
\end{figure}

Another way to look at jets is to consider back to back high transverse momentum hadron correlations. Fig.~\ref{f2} shows the measurements of such correlations for various collision types performed by the STAR experiment and reported in section~4.2 of Ref.~\refcite{Adams:2005dq}. Displayed are the azimuthal distributions of hadrons around a ``trigger'' particle of high enough $p_T$ to reflect the main direction of jets (4~GeV/c for the trigger particle and 2~GeV/c for the others in this example).
In p+p collisions (black histogram), one clearly sees particles belonging to both the narrower same ($\Delta \phi = 0$) and broader opposite ($\Delta \phi = \pi$) jets, while in central Au+Au collisions (blue stars) the away-side jet disappears\cite{Adler:2002tq}. This is also attributed to jet quenching, the away-side jet being absorbed by the dense matter produced at RHIC. As for the high $p_T$ suppression we saw in the previous section, this effect is not observed in d+Au collisions (red circles) where away-side hadrons are clearly distinguishable\cite{Adams:2003im}.

Jet-induced hadron production has been further and extensively investigated at RHIC and various effects corroborate the jet quenching hypothesis, among which:
\begin{itemize}
  \item In Au+Au collisions, the away-side disappearance grows with centrality. In fact, the most peripheral collisions exhibit a very similar away-side pattern as in p+p and d+Au collisions.
  \item The jets emitted in the reaction plane are less suppressed than in the perpendicular direction, where they have more matter to traverse\cite{Adams:2004wz}. In fact, the high $p_T$ (near-side) particles we see in central Au+Au collisions are likely to come from the periphery, the ``corona'', of the collision.
  \item By lowering the $p_T$ requirements (down to $\sim$1~GeV/c), one can find back the away-side jets\cite{Adams:2005ph}.
  \item These weakened away-side jets are depleted at $\Delta \phi = \pi$ and exhibit two displaced maxima around $\Delta \phi = \pi \pm 1.1$~radians\cite{Adler:2005ee}. This camel-back or conical-like shape provides insight in the quenched parton interactions with the medium. Various scenarios are proposed, such as radiative loss\cite{Polosa:2006hb}, \v{C}erenkov-like or Mach-cone emissions\cite{Ruppert:2005uz}. The later allows one to compute an average speed of sound in the medium of $c_S \sim 0.45$.
  \item Preliminary analyses of three particles correlations also exhibit the conical pattern\cite{Ulery:2006ha}. 
  \item The near-side jet exhibits a ``ridge'' along pseudorapidity (thus perpendicular to the azimuthal structure) that suggests the jets are indeed flowing with the expanding matter\cite{Adams:2005ph,Adams:2004pa}.
\end{itemize}

In brief, these high $p_T$ dihadron correlation studies show that \textbf{the matter is opaque} to jets to a first approximation, and clearly modifying their remaining structure.

\section{Elliptic Flow and Ideal Hydrodynamics}

Speaking of azimuthal correlation, it is noticeable that for moderate centralities, overlapping colliding nuclei form an almond-shape area. It is then relevant to look at the ``elliptic flow'' of particles, namely the second Fourier harmonic $v_2$ of the azimuthal distribution: $dN/d\phi = N_0 (1 + 2 v_1 \cos (\phi) + 2 v_2 \cos (2\phi) + \ldots).$ Experimentally, $v_2$ happens to be positive, meaning that the particle emission is enhanced in the plane of the reaction (along the smaller axis of the almond) with respect to the out-of-plane emission (along the larger axis). This reflects pressure gradients, i.e. strong interactions, that must exist at the very early stage of the collision to provide more transverse momentum to the emitted hadrons along the shortest axis. Moreover, the rather large values (up to $v_2\sim 20$~\% at 2~GeV/c) of the elliptic flow measured at RHIC contradict hadronic transport models (for instance accounting for only $\sim$~60~\% of the observed value\cite{Zhu:2005qa}). On the contrary, \emph{ideal} hydrodynamical models (for a list see section~3.5 of Ref.~\refcite{Adcox:2004mh}) that assume a QGP equation of state, a high energy density ($\epsilon \sim 20$~GeV/fm$^3$) and fast equilibration time ($\tau \sim$~0.6 to 1~fm/$c$) fits reasonably well a broad selection of data:
\begin{itemize}
  \item The transverse momentum dependence of elliptic flow is reproduced up to 2~GeV/c, and properly ordered for various species\cite{Adler:2003kt,Huovinen:2001cy} (from pions to cascades)\footnote{Being faster, higher $p_T$ particles share less the collective behavior of the bulk, which does not mean they do not see it, since we saw in the previous sections that they are very suppressed by this dense matter they traverse.}.  
  \item These $v_2(p_T)$ scales with the eccentricity ($\langle y^2 \rangle - \langle x^2 \rangle/\langle y^2 \rangle + \langle x^2 \rangle$) of the reaction for various collision systems, centralities and energies, underlining the facts that elliptic flow does reflect the very early stage of the reaction and that thermalization must arise rapidly\cite{Adare:2006ti}. 
  \item Hydrodynamics pressure gradients also imply a scaling by the transverse kinetic energy. While this property is verified for low $p_T$ (less than $\sim$~1~GeV/c) hadrons, it extends its validity to much higher $p_T$ when one divides both $v_2$ and $p_T$ by the number $n$ of constituent quarks\cite{Adare:2006ti}. This result holds for pions, kaons, protons, $\Lambda$, $\Xi$, but also for the $\phi$ mesons\cite{Abelev:2007rw,Afanasiev:2007tv} (of baryonic-level mass) and deuterons\cite{Afanasiev:2007tv} (with $n=6$), as shown on Fig.~\ref{f3}. 

\begin{figure}[htpb]
\centerline{\psfig{file=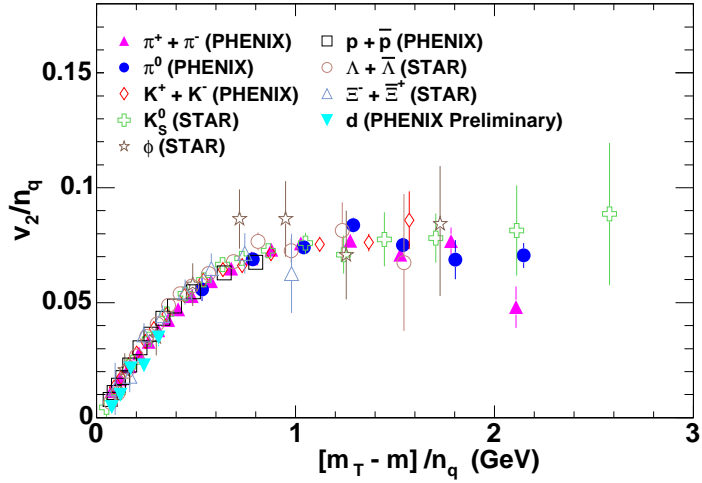,width=10cm}}
\vspace*{8pt}
\caption{Scaling of the elliptic flow parameter $v_2$ versus transverse momentum (left) or kinetic energy (right) for various particles. Both quantities are divided by the number of consistent quarks.  }
\label{f3}
\end{figure}

  \item The adjunction of even a low viscosity in hydrodynamical models deteriorates their fits to the data, in particular by moderating $v_2$ as $p_T$ grows\cite{Teaney:2003kp} (departing from ideal hydrodynamics around $p_T \sim 1$~GeV/c). The matter created at RHIC must then have a very low viscosity and was thus qualified as a ``perfect liquid''.
  \item The transverse mass spectra, i.e. the radial flow, are also reproduced by hydrodynamical models (with kinematic freeze-out temperature of $\sim$~100~MeV and transverse speed of $\langle \beta_T\rangle \sim 0.6$ for the most central Au+Au collisions\cite{Adams:2005dq}). 
\end{itemize}

This high degree of collective ideal hydrodynamical behaviors, setting up at very early times and exhibiting a low viscosity, tells us that \textbf{the matter is strongly interacting, in a liquid-like manner}.

\section{Baryons and Mesons}

\begin{figure}[htpb]
\centerline{\psfig{file=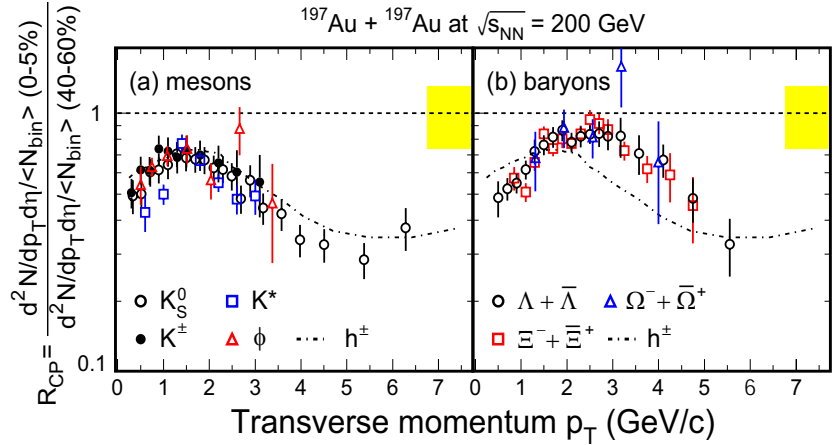,width=11cm}}
\vspace*{8pt}
\caption{Central to peripheral ratios of various mesons (left) and baryons (right) as a function of transverse momentum, as measured by the STAR experiment.}
\label{f4}
\end{figure}

We saw on Fig.~\ref{f3} that dividing by the number of constituent quarks helped the elliptic flow parameter $v_2$ to scale with transverse kinetic energy at moderate $p_T$ (2 to 4~GeV/c). This is not the first observable to exhibit a parton-like scaling.
Indeed, the nuclear modification factor $R_{AA}$ also shows a different pattern between baryons and mesons (including the $\phi$ which is of baryonic-level mass), in the same $p_T$ range.
Fig.~\ref{f4} from Ref.~\refcite{Adams:2005dq} shows the central to peripheral ratios (scaled by the number of collisions) of mesons (left) and baryons (right).
It is first noticeable that the relevant property to determine the fate of these intermediate $p_T$ particles is their baryonic/mesonic nature. Moreover, the fact that the baryon peak production is higher and lays at $\sim 3/2$ times the mesonic one, suggests that a quark coalescence or recombination mechanism is at play.

To test this hypothesis, the $p/\pi^+$ and $\overline{p}/\pi^-$ ratios can be studied in detail. Baseline p+p and peripheral Au+Au collisions exhibit very similar patterns, while the $p/\pi$ ratio is clearly enhanced in the moderate $p_T$ range. These particle ratios are equally reproduced by coalescence or recombination approach\cite{Abelev:2006jr}.

Added to the elliptic flow versus transverse kinetic energy scaling (Fig.~\ref{f3}), and to the partonic strength of jet quenching (Fig.~\ref{f1}), this result suggests that \textbf{the matter is of partonic nature}.

\section{Heavy Flavors Flow and Quenching}

\begin{figure}[htpb]
\centerline{\psfig{file=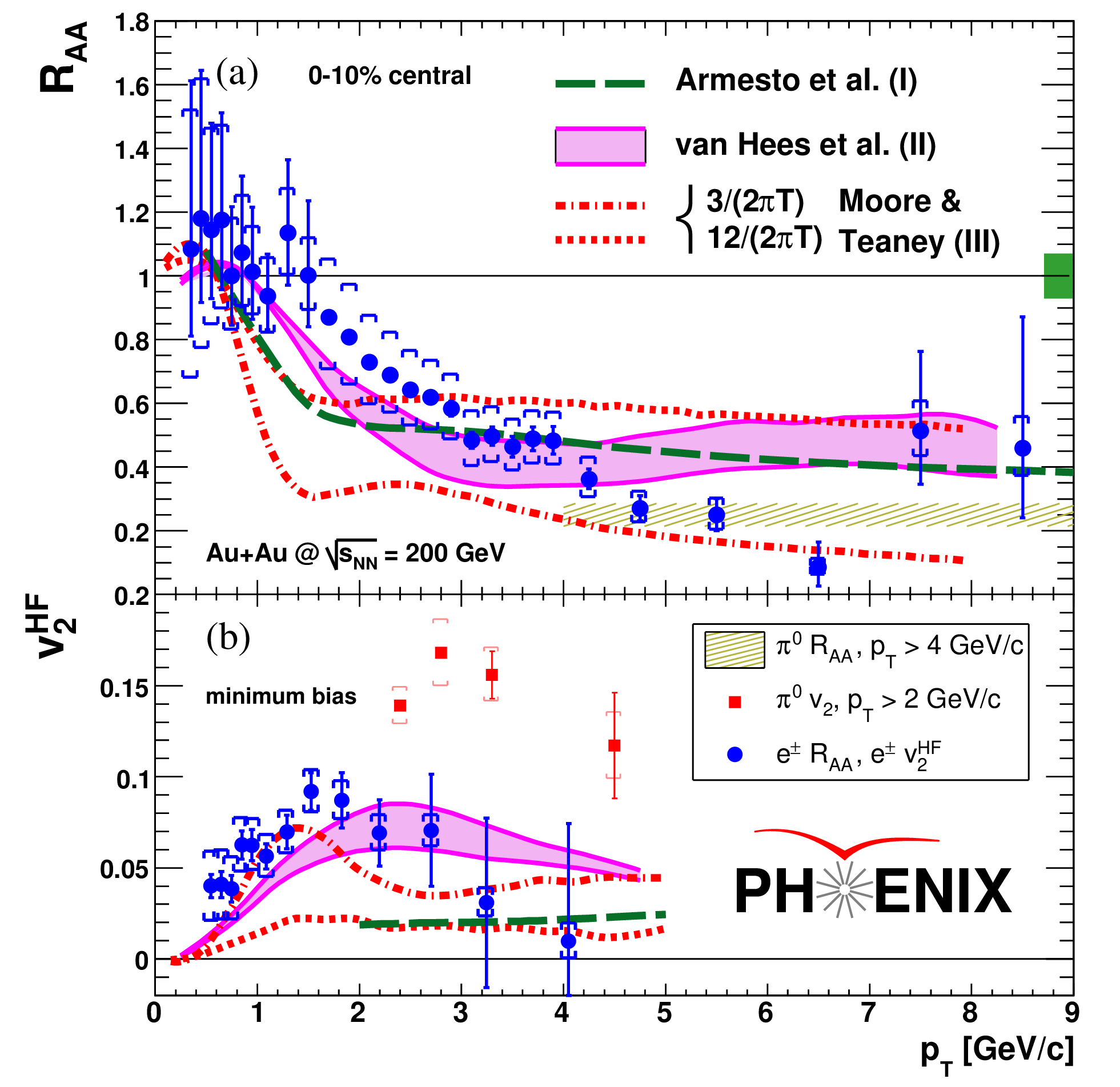,width=12cm}}
\vspace*{8pt}
\caption{Heavy flavor decay electrons in-medium behavior as measured by the PHENIX experiment, compared to $\pi^0$ and models. Top: Quenching in the most central Au+Au collisions. Bottom: Minimum bias elliptic flow. Both are as a function of transverse momentum.}
\label{f5}
\end{figure}

Being heavier, charm or bottom quarks are produced earlier than the light flavors, and their production yields can be in principle calculated by perturbative QCD. They are thus considered as good probes of the plasma earliest times. As we saw for light quarks, the nuclear modification factors and elliptic flow are good observables of the medium effects on produced particles. Fig~\ref{f5} shows both quantities for electrons from heavy flavor decays (blue circles) and $\pi^0$ (shaded band or red squares), as measured by the PHENIX experiment\cite{Adare:2006nq}. It is to be noted that even if the STAR\cite{Abelev:2006db} and PHENIX experiments disagree on the charm cross-section, they do agree on $v_2$ and $R_{AA}$. They both see that high~$p_T$ heavy quarks\footnote{The low~$p_T$ (up to 1.5~GeV/$c$) dominant yield scale with the number of collisions ($R_{AA} \simeq 1$) as expected.} are quenched by a factor of~5 and that they do exhibit a significant flow (up to 10~\%, while pions reach 20~\%). As for light flavors, both observables reveal a strong coupling to the medium.


These were surprises. Energy loss in a gluon medium was expected to be reduced for heavy quarks.
Indeed, in order to reproduce the data, one would need a much higher gluon density than the one required for light flavors ($dN_g/dy \sim 3500$ versus 1100, neglecting less quenched beauty decays\cite{Djordjevic:2005db}). Various hypotheses are made to reinforce the heavy quark quenching (adjunction of elastic energy loss, change in the charm/beauty ratio, modification of the strong coupling constant\ldots). Another approach is to quantify the medium effects by transport or diffusion coefficients. The models displayed on Fig.~\ref{f5} follow such approaches. With rather high values of these coefficients (\^{q}~=~14~GeV$^2$/fm in model I) they roughly manage to reproduce the amount of suppression and flow\cite{Armesto:2005mz,vanHees:2005wb,Moore:2004tg}.

Interestingly, and in order to illustrate the black hole/QGP connection mentioned as a foreword, the diffusion coefficient used in the models labeled II and III, corresponds to viscosity over entropy ratios that fall close to the lower quantum bound of $\hbar/4\pi k_B$ as derived through the AdS/CFT correspondence\cite{Kovtun:2004de}.

It is fair to say that the way the RHIC matter impacts heavy quarks is not perfectly understood yet, but it is also clear that it is strong. To that extend, I will dare to say that \textbf{the matter is ``tough''}, tough to understand as well as tough enough to shake the heavy flavors.

\section{Quarkonia Suppression}

\begin{figure}[htpb]
\centerline{\psfig{file=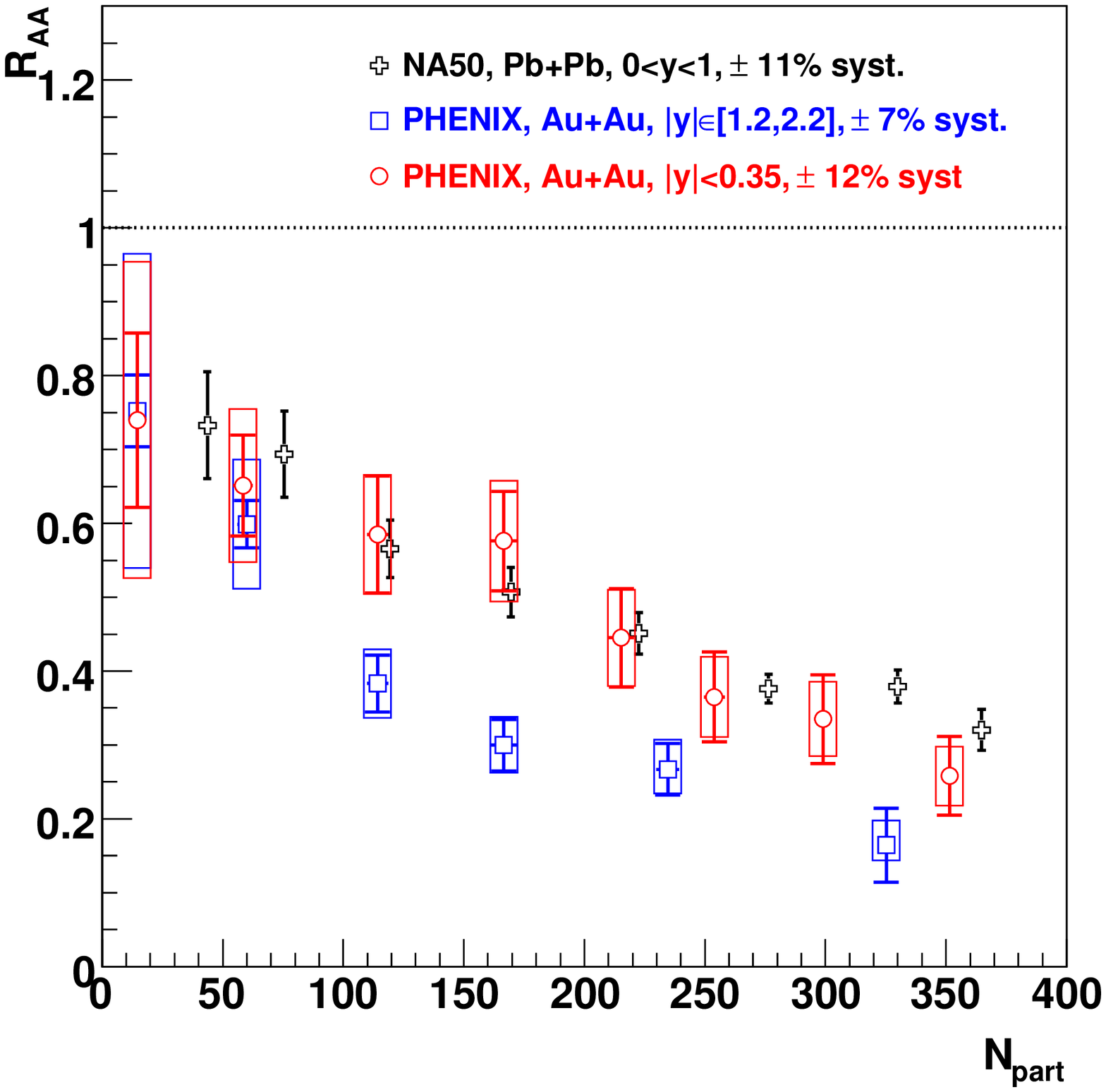,width=6cm} \psfig{file=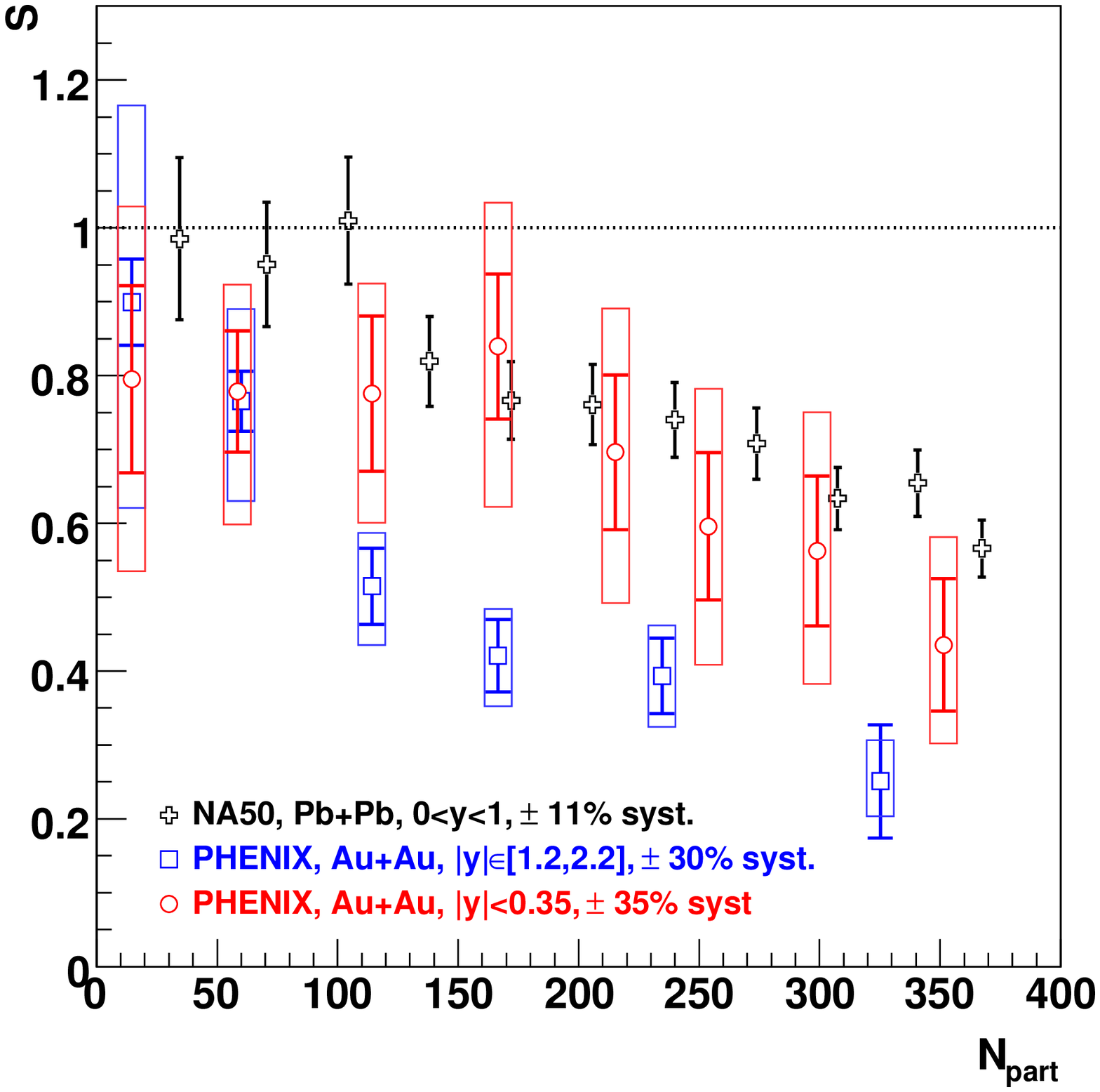,width=6cm}}
\vspace*{8pt}
\caption{$J/\psi$ suppression measured by the PHENIX and NA50 experiments, as a function of centrality, given by the number of participants. Left: nuclear modification factor. Right: $J/\psi$ survival probabilities after normal nuclear effects subtraction.}
\label{f6}
\end{figure}

We just saw that the bulk (low~$p_T$) charm production scales to first order with the number of binary collisions. This forms a good baseline for the study of bound states made of charm-anticharm quarks, the more stable of which being the $J/\psi$ particle. In fact, charmonia were predicted to melt in the QGP, due to Debye screening of the color charge\cite{Matsui:1986dk}. Furthermore, $J/\psi$ suppression was indeed observed at lower energy ($\sqrt{s_{NN}}= 17.3$~GeV) by the NA50 experiment\cite{Alessandro:2004ap} and is the main signature that led CERN to claim for the discovery of QGP. It was thus very awaited at RHIC energies. Fig.~\ref{f6} shows $J/\psi$ nuclear modification factors as measured by the PHENIX experiment\cite{Adare:2006ns}, for both midrapidity (red circles, $|y|<0.35$) and forward rapiditiy (blue squares, $1.2<|y|<2.2$), compared to the NA50 result (black crosses) and as a function of centrality (given here by the number of participants $N_{part}$). The midrapidity result is surprisingly similar to the pattern observed by the NA50 experiment, which also lies close to midrapidity ($0<y<1$). There is no fundamental reason for this to happen. First, the energy density for a given $N_{part}$ is much higher at RHIC and should further melt quarkonia.
Second, $J/\psi$ are known to be suppressed by regular nuclear matter as it is seen in p+A or d+A collisions\cite{Alessandro:2004ap,Adler:2005ph} and this \emph{normal} suppression should be different. In order to compare the two energy regimes, one first need to subtract the regular nuclear matter effects. Such an attempt\cite{GranierdeCassagnac:2007aj} is shown on the right part of Fig.~\ref{f6}. At RHIC, these effects are poorly constrained by a relatively low statistics d+Au dataset\cite{Adler:2005ph}$^{,}$\footnote{As well as by the fact that different p+p references were used for d+Au and Au+Au nuclear modification factors, which accounts for the $\sim$~30~\% global uncertainty quoted on the figure.} which significantly increases the uncertainty on the $J/\psi$ survival probability in the produced matter.

Anyway, we clearly see that $J/\psi$ are suppressed beyond normal nuclear effects, both at SPS and RHIC (especially at forward rapidity). Then, once these effects are subtracted, these facts remain:
\begin{itemize}
  \item SPS and midrapidity RHIC $J/\psi$ suppression are possibly still compatible. This lead to the hypothesis that direct $J/\psi$ do not melt neither at SPS nor at RHIC, but that only their feed-down less-bound contributors ($\chi_c$ and~$\psi'$) disappear in the QGP\cite{Karsch:2005nk}.
  \item $J/\psi$ are more suppressed at forward rapidity. This seems to contradict all models based on density-induced \emph{suppression}, in particular the original Debye screening hypothesis, as well as the sequential melting scenario of excited states suggested above.
\end{itemize}

Two ideas exist to explain this last surprising feature. First, gluon saturation could further suppress $J/\psi$ at forward rapidity, by playing a larger role than the one simply extrapolated from d+Au collisions (such an effect was computed for open charm in the CGC framework\cite{Tuchin:2004ue}).
Second, $J/\psi$ could be recreated in the plasma by recombination of independent charm and anticharm quarks (a large variety of recombination or coaelescence models\cite{Bratkovskaya:2004cq,Grandchamp:2004tn,Thews:2005fs,Yan:2006ve,Andronic:2006ky} exists). These two ideas do not provide quantitative predictions of the nuclear modification factors (recombination models suffering from the lack of input charm quark distributions). Other observables ($p_T$ and rapidity dependencies) are also available\cite{Adare:2006ns} but so far, they do not allow to rule out any possibility. New measurements are thus needed to further understand $J/\psi$ suppression at RHIC (higher statistics d+Au, $J/\psi$ elliptic flow, feed-down contributions...).

However, we do not need them to reckon that $J/\psi$ do melt beyond normal nuclear effects. This is a sign that \textbf{the matter is deconfining}.

\section{Thermal Radiation}

\begin{figure}[htpb]
\centerline{\psfig{file=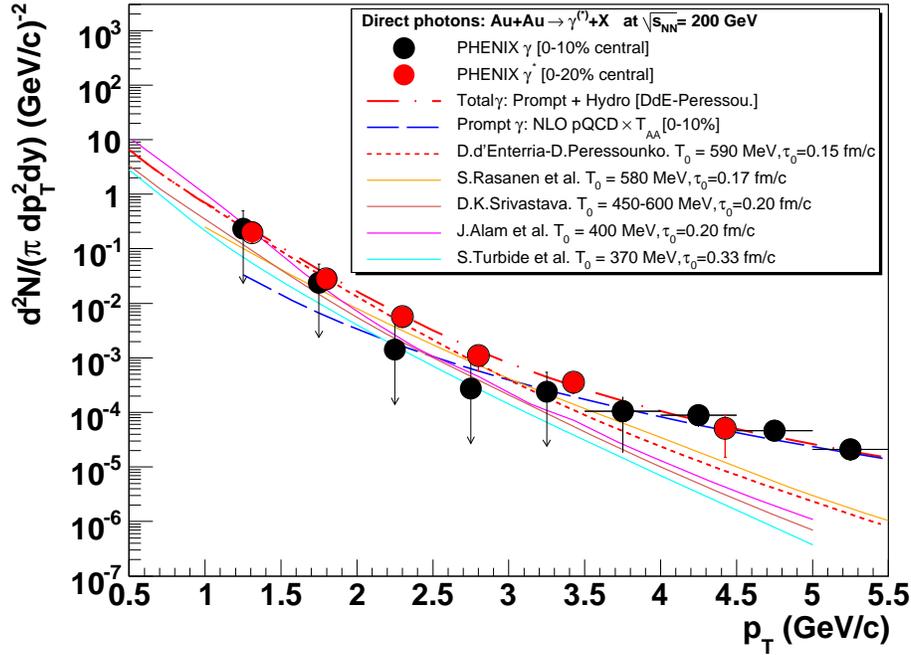,width=12cm}}
\vspace*{8pt}
\caption{Thermal + perturbative QCD fits to the photon yield in central collision, as seen by the PHENIX experiment (the internal conversion $\gamma^*$ yield is preliminary).}
\label{f7}
\end{figure}

Last but not least, a thermalized matter as the one suggested by the strong elliptic flow should emit its own thermal radiation. We saw on Fig.~\ref{f1} that photons are unmodified by the medium and the nuclear modification factor is compatible with unity. This holds for $p_T > 2$~GeV/c, but lower $p_T$ photons exhibit an enhancement when compared to perturbative, next-to-leading order QCD predictions\cite{Adler:2005ig}. This is further illustrated for the most central Au+Au collisions on Fig.~\ref{f7} on which the lowest $p_T$ photons\footnote{Note that the internal conversion $\gamma^*$ yield is a PHENIX preliminary result\cite{Peressounko:2007xe}.} deviate from the prompt contribution from a NLO pQCD calculation (dashed line). They are consistent with the addition of a thermal contribution. Various hydrodynamical models\cite{d'Enterria:2005vz} fairly reproduce the data assuming early (typically at a time of the order of 0.2~fm/$c$) temperature  of 400 to 600~MeV, well above the critical temperature of $T_c=190$~MeV provided by lattice QCD\cite{Karsch:2001cy} as the phase transition boundary to a quark-gluon plasma.

This result suffers the lack of \emph{experimental} p+p reference, but if NLO pQCD is taken as a baseline, we do see thermal photons that demonstrate that \textbf{the matter is hot}.

\section{Conclusions}

Even if we haven't (yet) observed any sharp change in the behavior of the Au+Au observables related to the predicted phase transition, nor numbered degrees of freedom, it is clear that the matter produced at RHIC behaves very differently than ordinary hadronic matter. Indeed, to answer the question raised by our title, we saw that the matter is gluon saturated, dense and opaque, strongly interacting and liquid-like, partonic and deconfining, ``tough'' and hot. It is thus very likely to be formed by deconfined quarks and gluons.

\section*{Acknowledgments}

It's my pleasure to thanks the CTP symposium organizers, in particular Daniel Den\'egri, for the invitation to speak in Cairo. I apologize to Peter Steinberg for having unintentionaly stolen his humoristic title\cite{Steinberg:2007iw}. My thanks also go to David d'Enterria, James Dunlop, Vi-Nham Tram and Fuqiang Wang for their corrections and comments.

%
%
%
%
%
%
%
%

\bibliographystyle{myunsrt}
\bibliography{Biblio}

\end{document}